\documentclass[twoside,twocolumn,english]{revtex4-1}
\usepackage[T1]{fontenc}
\usepackage[latin9]{inputenc}
\usepackage{amsmath}
\usepackage{graphicx}
\usepackage{esint}
\usepackage{xcolor}
\usepackage[normalem]{ulem}

\makeatletter
\usepackage{babel}

\makeatother

\usepackage{babel}
\begin{document}
\title{Argon, neon, and nitrogen impurity transport in the edge and SOL regions of a Tokamak}
\author{Shrish Raj, N Bisai, Vijay Shankar, and A Sen }
\affiliation{Institute for Plasma Research, A CI of Homi Bhabha National Institute,
Bhat, Gandhinagar, Gujarat, India-382428 }
\begin{abstract}
Numerical simulations of the interchange plasma turbulence in the
presence of medium-Z impurities (N$_{2}$, Ne, Ar) seeding have been
\textcolor{black}{performed} using BOUT++. These simulation results are used to study the
impurity transport mechanism in the edge and scrape-off layer (SOL)
regions. An analytical relation for the impurity
ion density with the vorticity, sources and sinks, and mass to charge ratio has \textcolor{black}{also} been derived. 
 Simulation shows that Ar$^+$  moves more strongly inward compared to N$^+$  and Ne$^+$ that has been verified 
 from the analytical relation. The most abundant species move both in the inward 
 and outward directions, but on average they mainly move outward. These behaviors
have been \textcolor{black}{confirmed} using cross-correlation techniques. The inward transport or negative flux of the impurity ions 
is found directly associated with the monopolar density holes in the presence of the electron temperature
gradient whereas the outward transport is associated with the plasma blobs.  The inward impurity transport
has been analyzed using hole fraction analysis. $\sim$44\% of Ar$^{+}$, $\sim$28\% of Ne$^+$, and
$\sim$25\% of N$^+$ ions of their total impurity densities are transported inward mainly through the avalanche events of the turbulent plasma.

\end{abstract}
\maketitle

\section{Introduction}

The anomalous plasma transport in the edge and scrape-off layer (SOL)
regions gives rise to large particle and heat loads on the plasma
facing material plates in tokamaks or stellarators. These loads can
damage the material plates and hence become a potential threat to
the safe operation of the machines. Therefore, these situations demand
a controlling mechanism so that these loads can be reduced \textcolor{black}{within} a safe operating limit. The medium-Z
impurities such as nitrogen (N${_{2}}$), neon (Ne), and argon (Ar) gases are found effective in controlling
the heat and particle fluxes in the boundary region of the tokamaks \cite{bisai2019dynamics,raj_effects_2020,raj2022studies,liu2018impact,jia2020simulations,J.Simpson_NME_2019_100599,Telesca_NME12_882,effenberg2017numerical}.
These gases and ions radiate and convert a major portion of energy
into radiation \cite{Kallenbach_2013}. The studies in references
\cite{liu2018impact,jia2020simulations,effenberg2017numerical} have
indicated about 40\%-50\% mitigation of the peak heat loads on the
divertor plates. A comparative study presented in ref.\cite{J.Simpson_NME_2019_100599}
suggests that neon is found to radiate more efficiently on closed
field lines than nitrogen. From ref.\cite{Telesca_NME12_882} it has
been found that if the seeding rate exceeds a certain threshold for a given particle 
and heat load crossing the separatrix then an inward pinch
is triggered which is mainly related to the reduction in the edge localized modes (ELM) activities.
Also, these impurities may play an important role to detach the plasma from the material
walls so that the direct heat and particle loads can be reduced on
these material plates; detailed detachment physics has been
reported in refs. \citep{krasheninnikov_2017,Kallenbach_2013,zhang2022modeling,coster2011detachment,O_Fevrier_PPCF_035017,Harrison_PPCF_065024,Liu_NF_2019_126046}.
\textcolor{black}{However} these impurity ions move inside the magnetic device and radiate,
hence, the inner plasma near the core cools down \cite{mollen2013impurity,pigarov2004multi,pigarov2005multi,priego2005anomalous}.
This is an undesirable situation - \textcolor{black}{could} be avoided if the inward
transport of the impurity ions is controlled. For this purpose, we
need to know the exact physical mechanisms for the inward transport
of the impurity ions.\\

The tokamak plasma has high levels of density fluctuations in the
edge and SOL regions mainly due to the interchange/drift wave instabilities
\citep{sarazin_intermittent_1998,BisaiPOP11P4018,DIppolitoPOP18P060501,GarciaPRL92P165003}.
These \textcolor{black}{unstable} modes make the plasma into a high turbulence states
that generate a type of coherent structure known as ``blob'' where
the local plasma density is a few times higher than the background density.
A generation of the plasma blob is also often associated with a generation
of plasma ``hole'' where the plasma density is lower than the background
plasma density. The blobs and holes move in the edge and SOL regions
and participate in the plasma transport \cite{krasheninnikov_2017,krasheninnikov2001scrape,BisaiPOP12P102515,BisaiPOP12P072520}.
\\

The plasma turbulence is modified by impurity seeding. In particular impurity ions which are relatively heavier compared to the main plasma ions contribute more strongly to the polarization drifts that modify the plasma vorticity and hence the plasma turbulence. The modification in the plasma vorticity again modifies the impurity ion dynamics and hence in this way the plasma-impurity interaction behaves self-consistently. Numerous studies
have been done experimentally \cite{Wang_NME_942,Barbui_NF59_076008,H.Tanaka_NME_2017_241}
as well as numerically in past related to the plasma-impurity interaction
in the boundary region of tokamaks. Authors have used a variety of numerical codes such as
SOLPS-ITER \citep{jia2020simulations,wang2022modeling}, SOLPS \citep{liu2018impact},
COREDIV \cite{Telesca_NME12_882}, EMC3-EIRENE \cite{effenberg2017numerical,KawamuraPPCF60P084005},
and EDGE2D-EIRENE \cite{J.Simpson_NME_2019_100599}. In the most recent works
\cite{raj_effects_2020,raj2022studies}, it has been found that
the impurity ions with single or double-positive charges due to the
plasma-impurity interactions move inward with a velocity of about
fraction of c${_{s}}$ (sound speed) so that these fluxes are negative.
The detailed investigations of plasma-impurity interaction shown in
Ref.\cite{krasheninnikov2008recent} suggest that the blobs not only
carry plasma but also carry fully stripped impurity ions from the
hot plasma towards the walls and the holes carry freshly ionized impurity
atoms from the walls towards the hotter region. In \textcolor{black}{R}efs.\cite{hasegawa2017impurity},
authors reported that the transport of impurity ions is associated
with the plasma blobs/holes \textcolor{black}{using 3D-PIC seeded blob/hole simulation. But transport in the presence of plasma turbulence associated with finite temperature gradient was missing.}
The present simulation work is dedicated to the study of impurity transport associated with these blobs and holes which are generated by the interchange plasma turbulence in the edge and SOL regions of tokamak in the presence of the finite electron temperature gradient that may differ from the dipolar distribution of the
impurity ions within the plasma blob/hole structures as described in Ref.\cite{hasegawa2017impurity}. \\
For this purpose, the impurity transport studies in the presence of
the nonlinear coupling between impurity and plasma turbulence have
been \textcolor{black}{performed} using two-dimensional (2D) numerical simulations. The main emphasis has been given to the transport mechanism associated with the propagation of blobs and holes where these are
generated self-consistently from the plasma turbulence. We have obtained
an analytical relation between the density of these primary impurity
ions and the vorticity in the presence of the impurity sources and
sinks as the minimum and maximum values of the vorticity normally correspond to the blobs and holes, respectively. 
The analytical relation has been validated with the numerical results.
It is found that the plasma vorticity is monopolar in the presence of finite electron temperature 
gradient \cite{BisaiPOP20P042509,shankar2021finite}. The first ionization species mainly
move inward in the edge region near the last closed flux (LCFS) and in the SOL region, all ions move
radially outward. The most abundant species in the edge region move inward and outward but
on average, they move radially outward direction. It is to be noted that the temperature in the hole is low such that recombination is preferred over ionization and therefore the higher charged species present in the hole will quickly recombine to the low ionization state. But in contrast, the temperature in the blob is high enough such that ionization dominates over recombination and hence the lower charged species will quickly ionize to the higher charged states.  To validate these facts we have presented the cross-correlation between the impurity ions and plasma density in the edge and SOL regions.
We have also quantified the amount of impurity transport by holes in the edge and SOL regions. It has been found in the simulation that the poloidal
and time-averaged density of impurity species related to plasma density holes are about 44\% of Ar$^{+}$, 28\% of Ne$^{+}$, and 25\% of N$^{+}$of their total impurity densities which are being transported radially inward in the edge region. \\

The paper is organized as follows. The numerical simulation method
has been described in section-\ref{sec:Model-Equations}. The derivation
of the analytical expression for the impurity ion density has been described in section-\ref{sec:Analytical-estimation}. 
The results related to the impurity transport mechanism from 2D simulations are described in section-\ref{sec:Simulation-Results}.
A brief summary and some concluding remarks are given in section-\ref{sec:Conclusion}.

\section{Model Equations and Numerical Simulation \label{sec:Model-Equations}}

The model equations used for the simulations couple the turbulence
in the edge and SOL regions with the medium-Z impurity gas dynamics.
Depending on the degree of electron impact ionization and recombination
rates, the impurity ions may exist in multiple charge states. The
model equations consist of the electron continuity equation, quasi-neutrality
relation, electron energy conservation, and conservation of the impurity
ions and the neutral gas. A simple diffusive model has been used to
describe impurity dynamics in the impurity conservation equation. We have assumed the ions of 
the main plasma and the ions of these impurities are cold. 
These equations have been described in refs.\cite{raj_effects_2020,raj2022studies} that have 
been used for this present simulation.
Numerically, all the radial derivatives have been computed using a
finite difference technique. The Neumann boundary condition (zero
gradient) has been applied to the electron density, electron temperature,
and density related to all ions in the radial direction. The poloidal
boundary conditions are periodic for all the variables, where Fourier
transformation has been used for all the variables in the poloidal
direction using the fastest Fourier transformation of the west version 3 (FFTW3) routines. The time evolution of all the species has been obtained using a fourth-order Adams-Bashforth method. The
simulation zone has a radial size $L_{x}=128\rho_{s}$ and a poloidal
size $L_{y}=256\rho_{s}$, where $\rho_{s}$ is ion gyroradius. We
have used 132 modes in the radial direction and 128 Fourier modes
in the poloidal direction. 132 modes in the radial direction with
$L_{x}=128\rho_{s}$ give a grid resolution of $\Delta x$ = 0.96$\rho_{s}$,
which is about $3.8\times10^{-4}$ m \cite{raj_effects_2020,raj2022studies}.
The Poisson brackets are solved using a third-order accurate up-winding
method. The potential has been solved from the vorticity using the
boundary value of the potential as $\Lambda T_{e}$ at the outer boundary (in SOL)
and $-T_{e}\ln(n)$ at the inner boundary (in edge), where $\Lambda$ is related
to the floating potential. The input parameters used in our simulations
are closely related to Aditya-U Tokamak parameters in the edge
and SOL regions and the other details of the model can be found in
refs. \cite{bisai2019dynamics,raj2020effect,raj2022studies}. The numerical
simulations have been done using BOUT++ framework \cite{DudsonPPCF53P054005}.

\section{Analytical estimation of the impurity ion density with vorticity
\label{sec:Analytical-estimation}}

In this section, we investigate a relationship related to the impurity ion density ($n_{imp}$)
with the plasma vorticity, and the impurity ion sources and sinks. The equation can be written as: 
\begin{equation}
\frac{\partial n_{imp}}{\partial t}+\vec{\nabla}_{\perp}\cdot(n_{imp}\vec{v}_{\perp imp})+\nabla_{\parallel}(n_{imp}v_{\parallel_{imp}})=S_{i}.\label{eq:impurity evolution}
\end{equation}
$\nabla_{\perp}$ and $\nabla_{\|}$ indicate the differential operators in ($x,y$) plane and $z$ direction. $t$ indicates time. $x$ and $y$
correspond to the radial and poloidal coordinates. The magnetic field $\vec{B}$ acts in $z$ direction. $\vec{v}_{\perp imp}$ is related
to $\vec{E}\times\vec{B}$ and polarization drifts that can be written as;

\noindent 
\[
\vec{v}_{\perp imp}=\frac{m_{imp}}{q_{imp}}\bigg(\frac{\partial\vec{E}}{\partial t}+(\vec{v}_{E}\cdot\vec{\nabla})\vec{E}\bigg)+\vec{v}_E,
\]
the first term is ion polarization drift and the second term is related to $\vec{E}\times\vec{B}$ drift; $\vec{v}_E=\vec{E}\times \vec{B}/B_0^2$. $\vec{E}$ represents the electric field, $B_0$ represents the magnitude of toroidal magnetic field. $\vec{v}_{\parallel imp}=\sqrt{\sum_{i}n_{imp}/n}c_{s,imp}$ where $c_{s,imp}$ is the common sound speed of ions \cite{BaalrudPOP18P023505}. $n$ indicates the electron density and $m_{imp}$ is the mass of the impurity. The ion diamagnetic term has not been taken into account because of the cold-ion approximation.  $S_{i}$ is the source and sink associated with the impurity ions obtained from the following chemical reactions for the impurity gas:\\

\begin{equation}
\left.\begin{aligned} & e^{-}+X^{j-1}\rightarrow X^{j}+2e^{-},\\
 & X^{j+1}+e^{-}\rightarrow X^{j},\\
 & X^{j}+e^{-}\rightarrow X^{j-1},\\
 & X^{j}+e^{-}\rightarrow X^{j+1}+2e^{-}.
\end{aligned}
\right\} \qquad\text{}\label{eq:reactions}
\end{equation}

\noindent $X^j$ indicates impurity ions, it can be any one of N$_{2}$, Ne, and Ar ions. $j$  indicates charge states. $j=1,2$,$\cdots$  indicate the first ionization, second ionization states, etc. The continuity equation of $X^{j}$ considering all the sources and sinks from the above reactions can be written as

\noindent 
\begin{equation}
\frac{dn_{imp}}{dt}-An_{imp}\frac{d\nabla_{\perp}^{2}\phi}{dt}-n_{imp}g\frac{\partial\phi}{\partial y}=S_{i}-\nabla_{\parallel}\cdot(n_{imp}v_{\parallel_{imp}}).\label{eq:normalized impurity equation}
\end{equation}
$A$ is the ratio of mass and charge of the impurity ions, i.e., $A=m_{imp}/(qj)$. $q$  indicates the charge of a proton. Here, $d/dt=\partial/\partial t+\vec{v}_{E}\cdot\vec{\nabla}_{\perp}$.
The loss of impurity ions in the parallel direction of the magnetic field in the sheath-connected region has been taken into account. Boussinesq approximation has been taken in the second term of Eq.(\ref{eq:normalized impurity equation}).
With these approximations, Eq.(\ref{eq:normalized impurity equation}) can be written as 
\begin{equation}
\frac{dn_{imp}}{dt}-n_{imp}\frac{d}{dt}\left[A\nabla_{\perp}^{2}\phi\right]+n_{imp}S_{l2}=S_{f2}+g\frac{\partial\phi}{\partial y}\label{eq:analytic}
\end{equation}
 $S_{f2}$ is the source of $X^{+}$ obtained from the first two reactions as indicated in Eq.(\ref{eq:reactions}). It represents the rate at which the impurity ions are being formed in the simulation. $S_{l2}$ is the effective sink related to the last two reactions as indicated in Eq.(\ref{eq:reactions}) including the parallel loss terms. It is to be noted that the dimension of $S_{l2}$ is per second so that $n_{imp}S_{l2}$ represents the loss rate of the impurity ions. Performing the time integration of Eq.(\ref{eq:analytic}) we can write the impurity ion density in Lagrange frame that moves with $\vec{E}\times\vec{B}$ drift velocity as: 
\begin{equation}
n_{imp}=S_{eff}\exp({A\nabla_{\perp}^{2}\phi})\label{eq:expression}
\end{equation}
where, 
\[
S_{eff}=e^{-\int{S_{l2}dt}}\left[\int({e^{-A\nabla_{\perp}^{2}\phi}e^{\int{S_{l2}dt}}S_{f2})dt}+c\right].
\]
$c$ represents the integration constant. The term $g(\partial\phi/\partial y)$ is not taken into account as  \textcolor{black}{$g=c_{s}^{2}/R$, where $R$ is the major radius, $c_{s}$ is the sound speed and its value in the dimensionless unit for ADITYA-U in the edge region is } small $g=6.3\times10^{-4}$.  
Equation-\ref{eq:expression} indicates that the impurity ion density is mainly governed by $A$, vorticity, and $S_{eff}$ terms. 
We have $A=m_{imp}/(qj)$, consequently the magnitude of  $n_{imp}$ will be lower for higher ionization species than the lower ionization species, which indicates that the magnitude of $n_{imp}$ for the higher ionization species drops exponentially for a given $S_{eff}$ and $\nabla_\perp ^2\phi$.  It is to be noted that in the absence of $S_{f2}$ and $S_{l2}$ Eq.(\ref{eq:expression})
corresponds to $S_{eff}=c$ that gives 
\[
n_{imp}=c\exp({A\nabla_{\perp}^{2}\phi}).
\]
Here, $c$ corresponds to the impurity ion density. This is the case
as derived by Hasegawa and Ishiguro in ref.\cite{hasegawa2017impurity}
where they showed that impurity ion density depends on the vorticity
of the plasma. A double peak of impurity ion density has been observed
within a plasma hole in their simulations {[}\citealp{hasegawa2017impurity}{]}
as the electron temperature was assumed uniform. But in our case, we have used finite $T_{e}$ gradient in
the model equation that may give a monopolar vorticity \cite{BisaiPOP20P042509,shankar2021finite}.
The presence of finite electron temperature in the edge region, the potential can be written as $\phi=\langle T_{e}\log(n)\rangle$.
As $T_{e}$ and $\log(n)$ will be maximum/minimum at the center of
the blob/hole, this will give a maximum/minimum of $\phi$ at the center
of a plasma blob/hole. The above relation indicates $\phi$ will be a function of the radius
(local) of the blobs and holes. Again, in the SOL region, $\phi$ is related with $T_{e}$ as $\phi=\Lambda \langle T_{e}\rangle $. The local radial electric
field ($E_{x}=-\partial\phi/\partial x$) of the blob/hole in the edge and SOL regions will give rise to spin/rotation in the direction of the local blob-axis (in the direction of the toroidal magnetic field) that will create a vortex structure.
Therefore, the blob/hole structures will be associated with the vortex.
Numerically, the validation of Eq.(\ref{eq:expression}) has been given in the Appendix using numerical data.

\section{Simulation Results \label{sec:Simulation-Results}}

\label{sec:results}

\begin{figure}
\begin{centering}
\includegraphics[width=8cm]{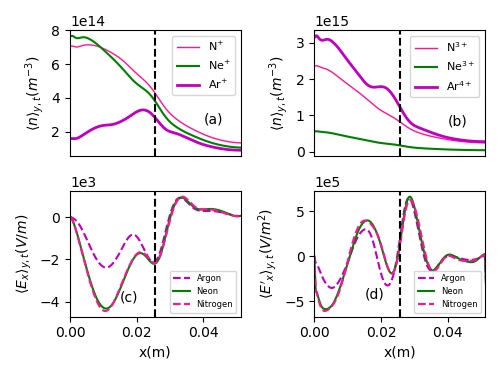} 
\par\end{centering}
\caption{\label{fig:Radial-equilibrium-of}(a) Radial distribution of singly
stripped impurity ion density (a), the impurity neutrals, and the abundant
ion species (b). (c) Radial variation of $E_{x}$. (d) Radial variation
of $E_{x}^{'}$. The vertical dotted line represents the position
of the last closed flux surface (LCFS) \textcolor{black}{and $\langle \rangle_{y,t}$ represents long time and poloidal average}.}
\end{figure}

\begin{figure}
\includegraphics[width=8cm]{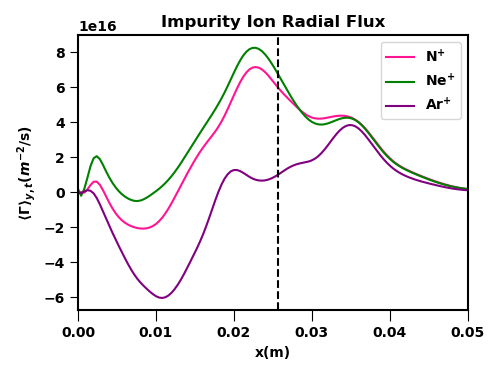}

\caption{The radial profiles of the impurity fluxes of N$^{+}$, Ne$^{+}$,
Ar$^{+}$ions obtained from the numerical simulation data using longtime
and poloidal averages. The position of LCFS has been shown by the
vertical dotted line. \label{fig:three_fluxes_sim}}

\end{figure}

Simulation data related to impurity transport will be presented here
in the statistically stationary phase of the turbulence. The equilibrium
profiles of the few impurity ions, radial electric field, and 
radial electric field shear are shown in Fig.\ref{fig:Radial-equilibrium-of}(a)-(d) as these
profiles play an important role in the transport processes.
The radial distribution of the first ionization species (N$^{+}$,
Ne$^{+}$, Ar$^{+}$) has been shown in Fig.\ref{fig:Radial-equilibrium-of}(a).
It shows that N$^{+}$, and Ne$^{+}$ ions densities are maximum in the
edge region near $x\sim0.01$m whereas Ar$^{+}$ ion density is maximum
near LCFS at $x\sim0.025$m. The ionization potential of Ar gas is
lower than N$_{2}$ and Ne gases so these gases are ionized at
the higher $T_{e}$ that is available at the lower values near $x=0.01$ m. Also the recombination
rate of the higher ionization such as Ar$^{2+}$ is higher than N$^{2+}$
and Ne$^{2+}$ is also playing a role in this behavior. The studies in
refs.\cite{raj_effects_2020,raj2022studies} indicate that in the
statistically stationary phase of the turbulence N$^{3+}$, Ne$^{3+}$,
and Ar$^{4+}$ ions are the most abundant. The radial profiles of
these ions are shown in Fig.\ref{fig:Radial-equilibrium-of}(b). These
profiles have been determined from various factors such as ionization
potential, recombination cross-sections, and the interaction of the
neutral gas with the turbulence as described in refs.\cite{bisai2019dynamics,raj_effects_2020,raj2022studies}.
The radial electric field $E_{x}$ has been shown in Fig.\ref{fig:Radial-equilibrium-of}(c).
It is found that $E_{x}$ is modified in the presence of Ar gas than N$_{2}$ and Ne gases mainly in the edge region. It is to be noted
that the modification in $E_{x}$ is related to the modification
in the plasma turbulence by plasma-impurity interaction. These impurities
are relatively heavier compared to the main plasma ions that contribute 
more strongly to the polarization drifts that modify the plasma vorticity
and hence the plasma turbulence. It is to be noted that the ionization/recombination
process alone cannot determine the radial profiles of the ions but
also the radial electric field shear $E_{x}^{\prime}$ plays an
important role in determining the radial profile of the impurity ions.
Therefore, we have plotted the radial profile of $E_{x}^{\prime}$
as shown in Fig.\ref{fig:Radial-equilibrium-of}(d). It has been found
that in the case of nitrogen and neon, $|E_{x}^{\prime}|$ is lower
than argon. The lower shear in the case of argon ions allows the inward/outward
motion more strongly than the nitrogen and neon ions. The radial profiles
of the impurity fluxes of N$^{+}$, Ne$^{+}$, Ar$^{+}$ions obtained
from the numerical simulation data have been shown in Fig.\ref{fig:three_fluxes_sim}. The flux of  Ar$^+$ is more negative
than Ne$^{+}$, N$^{+}$ions which indicates that Ar$^+$  mainly move in the inward direction than the other two ions. The fluxes of the other higher 
ionization species are mainly positive which are not shown here but we will discuss more in this section. \\

\begin{figure}
\centering{}\includegraphics[width=8cm]{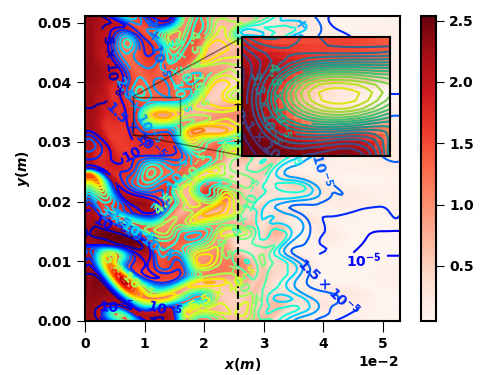}\caption{\label{fig:2D-contour_ar}Superposition of 2D contour plot of Ar$^{+}$
with plasma density indicated by colormap (``Reds''). Ar$^{+}$
is maximum at the density hole near LCFS. }
\end{figure}

\begin{figure}
\centering{}\includegraphics[width=8cm]{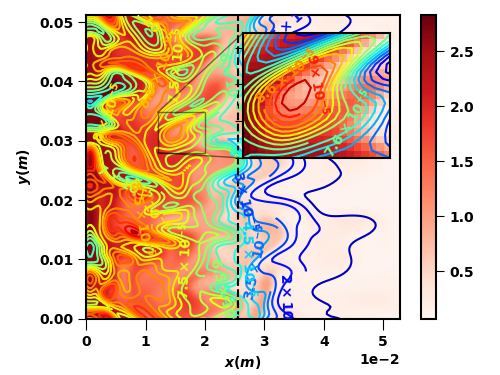}\caption{\label{fig:2D-contour_ne}Superposition of Ne$^{+}$ (contour lines)
with plasma density (``Reds'', colormap) . It indicates Ne$^{+}$
is maximum at the density hole near LCFS. }
\end{figure}

\begin{figure}
\begin{centering}
\includegraphics[width=8cm]{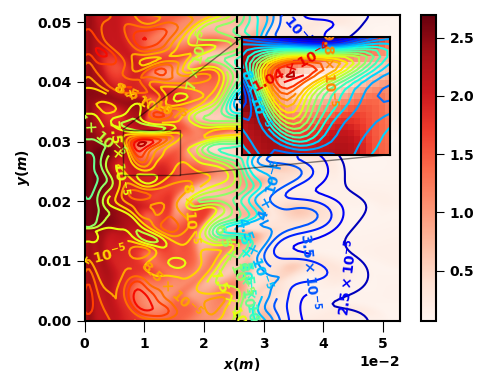} 
\par\end{centering}
\caption{\label{fig:2D-contour_ni}Superposition of contour plot of N$^{+}$
with plasma density indicated by colormap (``Reds''). This indicates
N$^{+}$ is the maximum at the density hole near LCFS. }
\end{figure}

Since the plasma particle transport is also associated with blobs
and holes \cite{krasheninnikov2008recent}, therefore, it is desirable
to analyze the impact of these impurity ions on these coherent structures
present in the plasma. As the present study is dedicated to the mechanism
of the impurity transport by the holes in the radially inward direction,
we superposed the first ionization species of argon impurity with
the plasma density in the $(x,y)$ plane as shown in Fig.\ref{fig:2D-contour_ar}.
The inset plot zooms a small rectangular area to show the presence
of Ar$^{+}$ inside a density hole. It is found that Ar$^{+}$ is
maximum near the center of the hole. A similar phenomenon is also seen
at the other radial positions near LCFS. In the SOL region, no such
phenomenon is seen. Similarly, the superposition of Ne$^{+}$ and
N$^{+}$ ion densities with the plasma density are shown in Fig.\ref{fig:2D-contour_ne}
and Fig.\ref{fig:2D-contour_ni}, respectively. These contour plots
show that for Ne$^{+}$ and N$^{+}$ density is maximum at a location
where the plasma density hole is present. In all three cases,
no such maxima of the impurity ions are seen within the plasma holes
in the SOL region. In the edge region, the heavier impurities have
modified the plasma vorticity more strongly, therefore, these impurities
behave in a more correlated way with the plasma vorticity that gives
the highest impurity density with the holes where the vorticity is
positive.\\

\begin{figure}
\begin{centering}
\includegraphics[width=8cm]{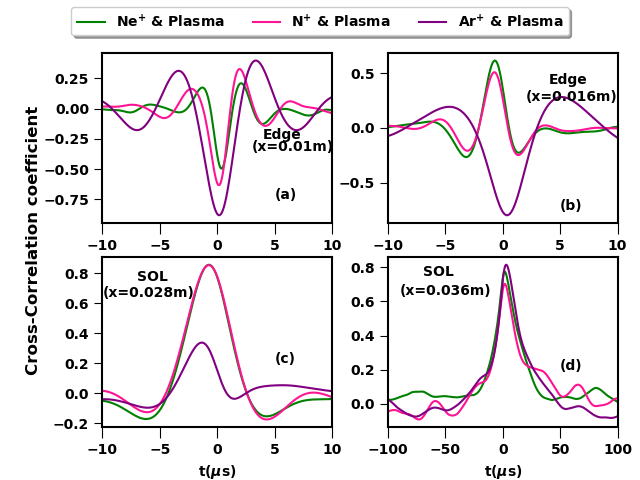} 
\par\end{centering}
\caption{\label{fig:corss_1st}Radial variation of plasma density and singly
stripped impurity ion density correlation. \label{fig:cross_1ions}}
\end{figure}

To show the simultaneous occurrence of impurity ion density maxima and plasma density holes 
for a large number of cases, we have presented cross-correlations between the two time series obtained
from the numerical simulations related
to the plasma density and the density related to the first ionization species at the four radial-poloidal 
(0.01m,0.025m), (0.016m,0.025m), (0.028m,0.025m), and (0.036m,0.025m) locations as shown in 
Fig.\ref{fig:corss_1st}. There is no definite choice of the four locations but these can help
to estimate the cross-correlation in other radial locations. The negative cross-correlation 
coefficient primarily in the edge region for the three cases indicates that when one density
increases, the other decreases. \textcolor{black}{The cross-correlation coefficient has been obtained using the relation: $C_r[\tau]=\sum_{t=0}^{T}s_1[t]*s_2[t-\tau]$, where $C_r$ is the cross-correlation coefficient for two different signals $s_1$ and $s_2$ obtained after taking the sum of their convolution. It is to be noted that $\tau$ varies from 0 to $T$}. The contour plots shown in Figs.\ref{fig:2D-contour_ar},
\ref{fig:2D-contour_ne}, and \ref{fig:2D-contour_ni} confirm that the impurity ion density is maximum 
where the plasma density is minimum (holes). Therefore, from the cross-correlation analysis and the contour
plots, we can conclude that the holes are mainly associated with the maxima of the first ionization 
species density. The first ionization species behave slightly differently than the main plasma, therefore,
the anti-correlation sustains up to a finite time, and after that they decorrelate. The simulation results indicate that 
when the holes move radially inward the impurity ions inside the holes also move inward
so that both are correlated till the cross-correlation time that is roughly defined by the time duration 
when the coefficient drops to $1/e$ of the maximum coefficient. This happens mainly near LCFS. \textcolor{black}{It is to be noted that the first ionization species inside the hole will be less 
probable to convert into second ionization species than the first ionization species outside the holes.
The holes maintain low temperature than the background up to the auto-correlation time, therefore, these will not be ionized further or less probable to convert into the second ionization species.}\\

The accumulation of the impurity ions with the density hole as shown in 
Figs.\ref{fig:2D-contour_ar}, \ref {fig:2D-contour_ne}, and \ref{fig:2D-contour_ni} could 
be explained using Eq.(\ref{eq:expression}), it
indicates that $n_{imp}$ will be maximum when $\nabla_{\perp}^{2}\phi$
is maximum. The interchange plasma turbulence indicates that $\nabla_{\perp}^{2}\phi$
is positive-maximum at the hole position and negative-minimum at the
blob position. Therefore, $n_{imp}$ will be maximum (minimum) at
the plasma hole (blob) location. This indicates that the impurity
ion density is found to be correlated with the positive vorticity
in the edge region or in other words the impurity ion density fluctuation
is mainly governed by instantaneous vorticity fluctuation. We have
described this behavior that is related to the first ionization species.
The same phenomenon is applicable to the second ionization species
but in this case $n_{imp}$ will be lower as it depends on $A$ where
$A$ will be half for the second ionization species than the first.
Similarly for the third, fourth, etc, species the impurity ion density
within a hole will be much lower. Fig.\ref{fig:cross_1ions}(b)
shows similar phenomenon as in Fig.\ref{fig:cross_1ions}(a), only
the cross-correlation coefficient is different. In the SOL region
at $x=0.036$m (Fig.\ref{fig:cross_1ions}(c)), and $x=0.028$m
(Fig.\ref{fig:cross_1ions}(d)), the cross-correlation is positive
which indicates that the impurity density increases with the decrease
of plasma vorticity in the time series. Therefore, one can conclude
that the impurity density is maximum at the position of the plasma blob.
This is quite different from the phenomena in the edge region. This
may be also because of the fact that both the impurity ions and blobs
move radially outward direction to facilitate the outward loss mechanism,
therefore, these are correlated. They are correlated till the cross-correlation
time as the impurity ions behave in a slightly different way than the main
plasma ions. It is to be noted that 2D simulation result indicates that in SOL region ($x=0.036$m), the
plasma holes are almost absent which indicates that the anti-correlation
with the impurity ions not seen. Figs.\ref{fig:cross_1ions}(a)-(d)
indicate that N$^{+}$ and Ne$^{+}$ behave differently than Ar$^{+}$
ions as Ar$^{+}$ has higher polarization drift than N$^{+}$ and
Ne$^{+}$ ions. Therefore, Ar$^{+}$ behaves more strongly with the plasma vorticity than the 
other two ions.\\

\begin{figure}
\begin{centering}
\includegraphics[width=8cm]{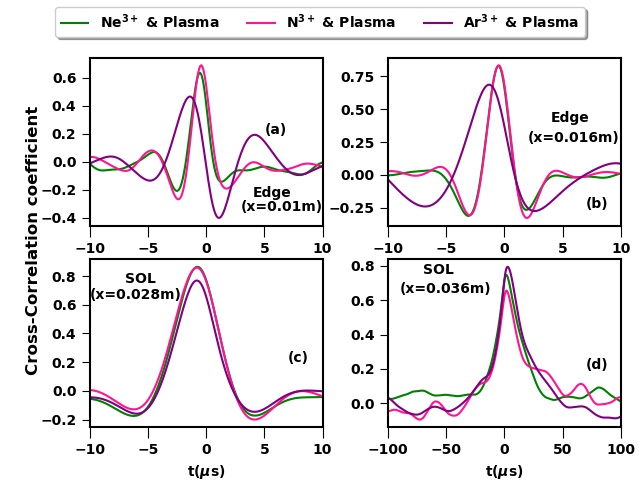} 
\par\end{centering}
\caption{\label{fig:cross_higher.png}Radial variation of plasma density and
the abundant ion species impurity ion density correlation}
\end{figure}

The cross-correlation between the plasma density and the most abundant
ion density of N$_{2}$, Ne, and Ar has been investigated as shown
in Figs.\ref{fig:cross_higher.png}(a)-(d). These are shown at the
same locations of Figs.\ref{fig:cross_1ions}(a)-(d). It is to be
noted that Figs.\ref{fig:cross_1ions}(a)-(b) and Figs.\ref{fig:cross_higher.png}(a)-(b)
show the opposite behavior. The correlation coefficient is negative
in Figs.\ref{fig:cross_1ions}(a)-(b) but in Figs.\ref{fig:cross_higher.png}(a)-(b)
these are positive. This behavior indicates that in the edge region,
the most abundant species move radially outward direction whereas
the first ionization species move inward. The high $T_{e}$
at lower $x$ ionizes the first ionization species further and produces
the higher ionization species. These higher ionization species accumulate
and the concentration builds up and when the gradient is higher than
the critical gradient (when interchange instability sets in) these
move radially outward due to the plasma turbulence. The blobs also
move radially outward direction, therefore, both the plasma blob and
the most abundant species move in the same direction which shows a
positive correlation coefficient. In the edge region, a small but
finite anti-correlation is observed as the holes in this region move
in the inward direction with the impurity ions as explained in the
previous paragraph.

The radial inward movement of the first ionization species can be
calculated from Eq.(\ref{eq:expression}) by estimating radial outward
flux as per

\begin{equation}
\Gamma_{imp}^{1+}=\left\langle n_{imp}\frac{E_{y}}{B}\right\rangle _{t,y}\sim\left\langle S_{eff}e^{A\nabla_{\perp}^{2}\phi}\frac{E_{y}}{B}\right\rangle _{t,y}.\label{eq:flux1}
\end{equation}

\begin{figure}
\centering \includegraphics[width=8cm]{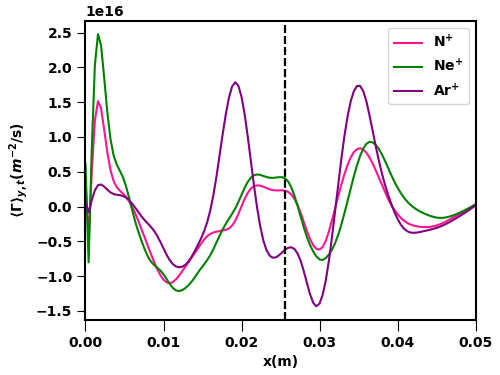}\caption{Comparison of the impurity ion flux using Eq.(5) where $\nabla_{\perp}^{2}\phi$
and $S_{eff}$ have been obtained from the numerical data. The vertical
dotted line indicates the position of LCFS.}
\label{fig:flux_com}
\end{figure}

\noindent Using numerical data we have estimated $\Gamma_{imp}^{1+}$,
the results have been shown in Fig.\ref{fig:flux_com}. The flux is
negative in the edge which indicates that these ions move inward direction.
The negative particle flux has been reported also in Refs.\cite{raj2022studies,raj2020effect}.
It is to be noted that Eq.(\ref{eq:flux1}) explains the negative
flux but the radial profiles as shown in Fig.\ref{eq:flux1} are different
from the actual numerical results as shown in Fig.\ref{fig:three_fluxes_sim}.
This mismatch is mainly because we have estimated the
flux in the Lagrange frame but the numerical results from the simulation
have been done in the laboratory frame. 

\begin{figure}
\begin{centering}
\includegraphics[width=8cm]{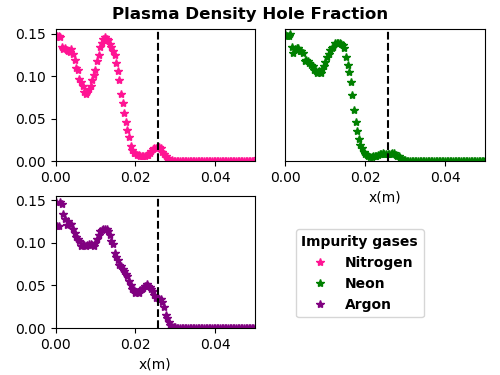}
\par\end{centering}
\caption{\label{fig:Hole-fraction}The plasma hole fraction in the presence
of N$_{2}$, Ne, and Ar gases. The vertical dotted lines indicate
the position of LCFS.}
\end{figure}

\begin{figure}
\begin{centering}
\includegraphics[width=8cm]{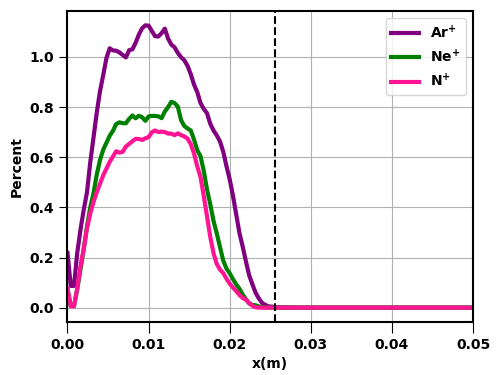}
\par\end{centering}
\caption{\label{fig:Fraction-of-impurity} The fraction of impurity ions within
the holes. The dotted line indicates the position of LCFS.}
\end{figure}

As the holes are playing a role in the impurity ion transport mainly
for the first ionization species, we have investigated the hole fraction in the presence of impurity gases.
The hole fraction is defined as the fraction of plasma that is contained
within a hole. The low hole fraction indicates fewer numbers of plasma
electrons with electron temperatures lower than the background. The
radial distribution of the fraction of the plasma density in the holes
is shown in Fig.\ref{fig:Hole-fraction}. The radial distribution
of the hole fraction suggests that the holes are mainly present in the
edge region and there are almost no holes in the SOL region. The low
electron temperature in the holes will provide a favorable condition
for the inward impurity transport as the low electron temperature
will lower the electron impact ionization events of the first ionization
species. Therefore, the first ionization species will survive within
the hole for a long time and will move much inside before reaching
a high-temperature zone where the hole will be destroyed or the impurity
will be ionized by the high temperate in that zone. In the case of
Ar$^{+}$ions (unlike N$^{+}$ and Ne$^{+}$), the negative gradient
provides stability against the interchange instability that will also
favor the existence of Ar$^{+}$ ions at the lower $x$, therefore,
large numbers of Ar$^{+}$ions move in the inward direction to provide
large negative impurity flux by $\vec{E}_{y}\times\vec{B}$ drift.
$E_{y}\times B$ drift will be the same for both the impurity ions
and main plasma particles so the structure as a whole convects in
the radially inward direction. We have also estimated the percentage
of impurity transport associated with the plasma density holes for
all three gases from the numerical data shown in Fig.\ref{fig:Fraction-of-impurity}.
The conditional integration of the poloidal average of the impurity
ion density is the quantification technique used in the present simulation
and is represented by Eq.(\ref{eq: Trapping fraction})

\begin{equation}
\textcolor{black}{f_{h}}=\frac{\intop_{0}^{L_{x}}Hn_{imp}dx}{\intop_{0}^{L_{x}}n_{imp}dx},\label{eq: Trapping fraction}
\end{equation}
where $H$ is the conditional parameter. It yields, $H=1$ for an
event when the holes and maxima of the impurity ions exist simultaneously,
otherwise $H=0$. The quantity \textcolor{black}{$f_{h}$} gives the impurity fraction associated
with the density holes. 

The percentage of first ionization species associated with the holes
has been shown in Fig.\ref{fig:Fraction-of-impurity}. The area under
each curve represents the total amount of impurity ions associated
with the density holes. We have calculated the area from the simulation
data which gives $\sim$44\% of Ar$^{+}$, $\sim$28\% of Ne, and
$\sim$25\% of N ions that are transported inside the edge region
through the turbulent events.

\section{Conclusion and discussion \label{sec:Conclusion}}

Numerical simulations have \textcolor{black}{performed} to show the modification of the plasma turbulence
by the medium-Z (N$_2$, Ne, and Ar) impurity gases in the edge and SOL regions. The main emphasis 
has been given to the dynamics of the impurity ions. It is found that these impurity ion motions
are strongly governed by vorticity. We have derived an analytical relation for the impurity
ion density with vorticity, sources and sinks, and mass-to-charge ratio. Ar$^+$  moves more strongly
inward compared to N$^+$  and Ne$^+$. The most abundant species move both in the inward and outward directions, but
on an average they mainly move outward in the edge and SOL regions. We have quantified this behavior
using cross-correlation techniques. The numerical simulations indicate that the impurity ions' inward transport (negative flux) is directly associated with the monopolar density holes in the presence of the electron temperature
gradient. In contrast, outward transport is associated with plasma blobs.  The results presented here might 
help in the future to understand/optimize the radiative loss in tokamak-related experiments. The inward impurity transport
has been quantified using hole fraction analysis. It is found that  $\sim$44\% of Ar$^{+}$, $\sim$28\% of Ne$^+$, and
$\sim$25\% of N$^+$ ions of their total impurity density that are transported inside the edge region through the intermittent events of the turbulent plasma.

\section{Acknowledgements \label{sec:Acknowledgement}}
The simulations are performed on the Antya cluster at the Institute for Plasma Research (IPR), India. A.S. is thankful to the Indian National Science Academy (INSA) for its support under the INSA Senior Scientist Fellowship scheme.

\section{Appendix}

\begin{figure}
\centering \includegraphics[width=8cm]{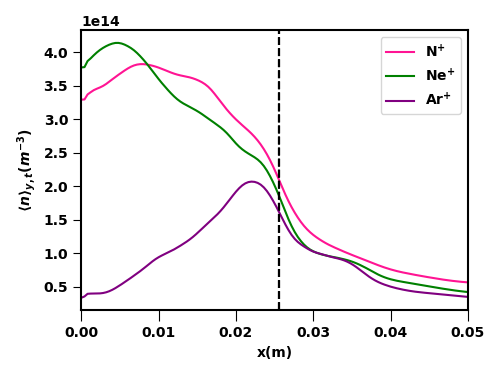}\caption{The radial profiles of $n_{imp}$ as obtained from Eq.(\ref{eq:expression})
and from the numerical simulation data. The vertical dotted line shows
the position of LCFS. \label{fig:analytical_den}}
\end{figure}

In order to verify the theoretical estimation of Eq.(\ref{eq:expression}),
we have calculated $S_{eff}$ and $\nabla_{\perp}^{2}\phi$ from the
numerical data as described in Section-\ref{sec:Simulation-Results}
and have estimated $n_{imp}$. Long time and poloidal averages of $n_{imp}$
have been shown in Figure-\ref{fig:analytical_den}. These results
are nearly comparable with Fig.\ref{fig:Radial-equilibrium-of}(a)
that differ by a factor of two mainly because of the nonlinear terms
but the radial variations of the profiles are almost similar. These
results justify the Eq.(\ref{eq:expression}).

\noindent 

\bibliographystyle{unsrt}
\bibliography{bibdata}

\end{document}